\documentclass[12pt,a4paper,oneside]{article}
\usepackage[german, english]{babel}
\usepackage{amsmath} 
\usepackage{mathtools} 
\usepackage{amssymb} 
\usepackage{upgreek}
\usepackage{graphicx}
\usepackage{epstopdf}
\usepackage{wrapfig}
\usepackage{nicefrac} % for fractions within text.
\usepackage{capt-of}
\usepackage[font=small]{caption}
\usepackage{subcaption}
\usepackage{hyperref}
\usepackage[backend=biber, hyperref=true, url=false, doi=false, isbn=false, style=numeric-comp,
            block=none, maxbibnames=2,firstinits=true, eprint=false, sorting=none]{biblatex}
\usepackage[left=28mm, right=21mm, top=21mm, bottom=21mm]{geometry}
\usepackage[utf8]{inputenc}
\usepackage{csquotes}
\usepackage{color}
\usepackage{lscape}
\usepackage{bm}
\usepackage{bbold}
\usepackage{parskip}
\usepackage{setspace}
\usepackage{titlesec, blindtext, color}
\definecolor{gray75}{gray}{0.75}
\newcommand{\hsp}{\hspace{20pt}}
\titleformat{\chapter}[hang]{\Huge\bfseries}{\thechapter\hsp\textcolor{gray75}{$|$}\hsp}{0pt}{\Huge\bfseries}
\author{Tobias Marx}
\date{\today }
\title{Mapped infinite elements in practice}
\begin{document}
%\indent
%\pagestyle{plain}
\setlength\parindent{6mm}
\setlength{\parskip}{2mm}
\renewcommand{\vec}{\bm}
\renewcommand{\i}{\mathrm{i}\mkern1mu}
\newcommand{\mat}[1]{\mathbb{#1}}
\newcommand{\prog}[1]{\texttt{#1}}
\newcommand{\eq}[1]{eq.~(\ref{#1})}
\renewcommand{\sec}[1]{sec.~(\ref{#1})}
\newcommand{\rt}[1]{\textcolor{red}{#1}}
\newcommand{\bt}[1]{\textcolor{blue}{#1}}
\newcommand{\eqs}[1]{eqs.~(\ref{#1})}
\newcommand{\fig}[1]{fig.~(\ref{#1})}
\newcommand{\refs}{\rt{[refs]}}
\newcommand*\colvec[3]{
    \begin{pmatrix}#1\\#2\\#3\end{pmatrix}
}
\widowpenalty=100000 %avoid hurenkind    ->
\clubpenalty=100000 %avoid Schusterjunge ->
\displaywidowpenalty=100000
\maketitle

%\section{Introduction}
\paragraph{Disclaimer:}
This document is written mainly as a documentation and reference for the implementation of infinite elements in the finite element library libMesh~\cite{Kirk_EC_2006} and aims to be a concise description of the technical details also providing background and a brief derivation of the expressions.

To the author's knowledge, the library libMesh has the only open-source implementation of infinite elements.
%Initially, this used an iterative approach for computing quantities on the reference elements.
The present article discusses the algorithms implemented in libMesh to compute the required quantities for assembling the system matrices.
The speciality is that it uses analytic expressions based on the physical element instead of the reference element, which is not only faster but also numerically more accurate than the previously used iterative approach.
The key advantage of the protocol is that it allows computing quantities for the limit $r\rightarrow \infty$, which is not possible in most other approaches.

\section{Introduction}
Infinite elements represent a large class of special-purpose elements used to amend finite element simulations that treat problems in unbounded domains.
The present article focuses on elements designed for solutions of the Helmholtz equation
\begin{equation*}
 \Delta \phi +k^2 \phi=0
\end{equation*}
for $k \in \mathbb{R}$ and with outgoing  boundary conditions (also referred to as Sommerfeld radiation condition)~\cite{Sommerfeld_AP_1949}
$$\lim_{r\rightarrow \infty} r(\frac{\partial}{\partial r} \phi(\vec{r})-\i k \phi(\vec{r}))=0$$ 
where $\vec{r}=(x,y,z)$ is a spatial vector and $r=\sqrt{x^2+y^2+z^2}$ the radial coordinate associated with it.

%Since the infinite elements developed for different applications differ in their properties, the discussion is largely limited to those developed for the Helmholtz equation and which have only one infinite direction using a spherical mapping \rt{REFS}.
Even though elements for various geometries were developed~\cite{Schoder_JCP_2019,Toselli_CMAME_1997}, here only mapped type infinite elements with a spherical mapping scheme are considered, as they are the most versatile group.
However, most of the results can be generalized to spheroidal mappings, while formulations with several infinite directions or elements with a Cartesian mapping (\textit{i.e.} whose rays are parallel) have different properties and must be considered separately.
%\rt{check the terms spherical mapping, Cartesian mapping}

Being the most popular type of mapping, the focus in this document is on the Zienkiewicz mapping, but the results hold at least qualitatively also for more general forms.
Since the technical aspects of the mapping often lack details in the literature, they are discussed briefly in~\sec{sec:map}.
In~\sec{sec:problem}, the divergence of the mappings' Jacobian with increasing radius is discussed.
While this does not challenge the numerical accuracy in most cases, changes in the matrix setup as presented in~\sec{sec:matrix} are advantageous.
%To obtain a stable and reliable integration of matrix elements, some aspects should be adapted for infinite elements because (as discussed in~\sec{sec:problem}) the Jacobian of infinite elements diverges. , requiring changes 
%While most quantities can be computed using the usual Lagrange interpolation scheme, the complex phase $\mu(\vec{r})$ cannot be described easily on the reference element, and thus better is computed on the physical element.
%
%The most prominent forms of these are denoted by Astley-Leis elements\rt{REFS}, Bettess elements\rt{REFS}, Zienkiewicz \rt{REFS}, 
%The main aspect here is presented in~\sec{sec:problem} and~\sec{sec:matrix}, discussing the divengence of the Jacobian mapping the reference element to the physical space and the setup of the system matrix in this context.

%\subsection{The General form of Zienkiewicz Mapping}
\section{The mapping from the Reference Element}
\label{sec:map}

Most modern infinite elements use a mapping of a finite reference element to the infinitely large physical element.
%Commonly, a mapping from $[-1:1)$ to $[a:\infty)$ is used, and thus the reference element can be a common prism-shaped or hexahedral element.
It is common to map the radial direction from $[-1:1)$ to $[a:\infty)$ allowing the reference element to be a standard prism-shaped or hexahedral element.
%It should be kept in mind during the following discussion that the inner radius $a$ varies across the element and hence has a non-zero derivative.
Since the inner radius $a$ varies across the element, it also is a variable with a non-zero derivative.
The mapping has the form 
%Many works refer \rt{refs} to the mapping using the notation 
\begin{equation} \label{eq:gen_map}
  r= \sum_i^N M_i r_i 
\end{equation}
suggested by Marques and Owens~\cite{marques_CS_1984}.
%which is the form referred to by many authors~\cite{angelov_infinite_1991,Astley_JSV_1994}\rt{refs}.
%by Marques and Owens~\cite{marques_CS_1984}.
In practice, one usually chooses $N=2$ with $r_1=a$ and $r_2=2a$ where $a$ is the distance of the elements' base face to the origin and $M_1=-\frac{2\zeta}{1-\zeta}$, $M_2=\frac{1+\zeta}{1-\zeta}$ are functions of the reference space coordinate $\zeta$.
While this is a useful choice, the above form is a nontransparent way of writing $r=\frac{2a}{1-\zeta}$ which was suggested earlier by Zienkiewicz~\cite{Zienkiewicz_IJNME_1983} and can be accomplished with $N=1$.

A description for the construction of more general schemes of the form \eq{eq:gen_map} is given, \textit{e.g.}, by Bettess ~\cite[ch. 4.3]{Bettess_1992}.
Analysis of this scheme shows that, in the general form, the resulting mapping function is $r=\frac{p_n(\zeta)}{1-\zeta}$ where $p_n$ is an $n$-th order polynomial in $\zeta$ which should be free of nodes to give an invertible mapping.

\section{The problem: Diverging Jacobian}
\label{sec:problem}
\begin{wrapfigure}{r}{.5\textwidth}
  \includegraphics[width=.5\textwidth]{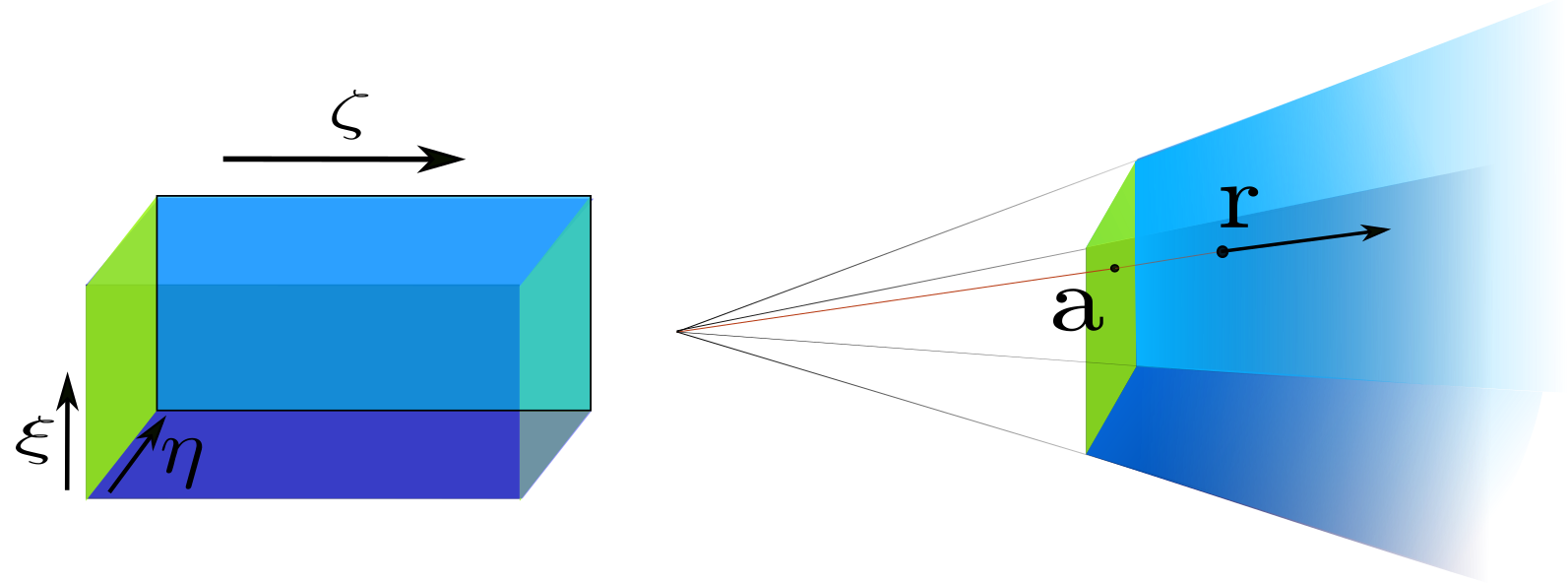}
  \caption{Schematic representation of the mapping of a hexahedral element: the direction $\zeta$ is mapped to an infinite range}
  \label{fig:inf-elem}
\end{wrapfigure}
%While the divergence of the Jacobian is clear from the infinite extend of the element, the reason for the dependence on $d^2=r^{\rm{dim}+1}$ may be not very clear on the first glance.
While the Jacobian of a finite to an infinite mapping must be divergent, it is not immediately clear that the Jacobian of infinite  elements diverges proportional to $r^{\rm{dim}+1}$, where dim is the spacial dimension.
Before looking at the formulae in the next section, a qualitative discussion helps to understand the origin of this dependence.
%While the infinite extend of the element clearly implies that the Jacobian must be divergent, it is not immediately clear that it should diverge to the order $d^2=r^{\rm{dim}+1}$ may be not very clear on the first glance.

The natural quantity to look at is the volume of the element, $\int_R dV$, truncated at radius $R$. 
In physical space it extends not only infinitely in radial direction, but also its length in the other two directions grows linearly with increasing radius, so that the volume diverges with $R^{\rm{dim}}$.
Transforming this integral into reference space gives $\int_R dV = \int_R J d\xi d\eta d\zeta \propto r^{\rm{dim}}$ with the Jacobian $J=J(\xi,\eta,\zeta)$ of the correspnding transformation.
Since, as discussed above, the mapping is (at least asymptotically) $r \propto \frac{1}{1-\zeta}$, one can write $\int_R J d\xi d\eta d\zeta \propto \zeta^{-\rm{dim}}$ and thus the integrand $J\propto \zeta ^{-(\rm{dim}+1)}$ which, transformed back to the physical space, gives $J\propto r^{\rm{dim}+1}$.
%This holds everywhere for Zienkiewicz mapping and asymptotically for any other mappping of the Marques and Owens form.

%In practice, this divergent behavior must be circumvented by modifying the procedure of matrix assembly, \textit{e.g.} as discussed in section \ref{sec:matrix}.
In practice, the Jacobian is only evaluated at finite distances, but if high-order integrations are needed, the growing Jacobian may lead to numerical inaccuracies.
However, this can be circumvented by modifying the procedure of matrix assembly as discussed in section \ref{sec:matrix}.

\subsection{The Jacobian for the Zienkiewicz mapping}
While for practical purposes the Jacobian for the mapping of the physical space to the reference space is relevant, its inverse is easier to obtain because it has a tensorial structure.
Thus, the mapping of the reference space to physical space is considered, and is inverted afterwards.

The reference space coordinates of a given point $\vec{r}$ are 
\begin{align*}
  \xi(\vec{r}) &= \xi(\vec{a}) \\
  \eta(\vec{r}) &= \eta(\vec{a})\\
  \zeta(\vec{r}) &= 1-\frac{2a}{r} 
\end{align*}
%where $\vec{a}=\frac{\vec{r_0}\cdot \vec{n}}{\vec{r}\cdot \vec{n}} \vec{r}$ is the projection of $\vec{r}$ onto the base element and $a=|\vec{a}|$.
%where $\vec{a}$ is the projection of $\vec{r}$ onto the base element and $a=|\vec{a}|$.
where $\vec{a}$ is radius vector of the crossing point with the base element (left face of the element in \fig{fig:inf-elem}) and $a=|\vec{a}|$.
%The vector $\vec{n}$ is the base elements normal and $\vec{r_0}$ is an arbitrary point on the plane. For non-planar elements (non-affine mappings), $\vec{n}$ and $\vec{r_0}$ are functions of $\vec{r}$.

In the following, the mapping of $\xi$ and $\eta$ is assumed to be known at the base of the element since it corresponds to the mapping of an outer face of the neighboring finite element.
The corresponding base coordinate is denoted as $\vec{r}_{\rm{base}}$.
While formally it coincides with $\vec{a}$, in the following $\vec{r}_{\rm{base}}$ is used to distinguish the dim$-1$-dimensional subspace of the base face from the full-dimensional space.
Moreover, the expression
\begin{equation}\label{eq:rbase}
   \vec{r}_{\rm{base}}=a \hat{\vec r} %\frac{\vec{r}}{r}
\end{equation}
with the radial unit vector $\hat{\vec r}=\frac{\vec{r}}{r}$ is useful to locate the base point corresponding to $\vec{r}$.

%Since this coincides with the outer face of the finite element, this assumption should be easy to fulfill.

%For the computation of the Jacobian the Zienkiewicz mapping $r=\frac{2a}{1-\zeta}$ is used where $r=\sqrt{x^2+y^2+z^2}$ is the radial coordinate and $a$ corresponds to the distance $r_{\rm{base}}$ of the corresponding base point to the origin.

Using the above equations, the mapping of the reference space $(\xi$, $\eta$, $\zeta)$ to $(x$,$y$,$z)$ reads
\begin{align*}
   %\frac{\partial \xi}{\partial \vec{r}} &= \frac{\partial \xi}{\partial \vec{r}_{\rm{base}}} \frac{a}{r} 
   %                     + \left( \nabla a - \frac{a\vec{r}}{r^2} \right)\left( \frac{\vec{r}}{r} \cdot \nabla \frac{\partial \xi}{\partial \vec{r}_{\rm{base}}} \right)\\
   %\frac{\partial \eta}{\partial \vec{r}} &= \frac{\partial \eta}{\partial \vec{r}_{\rm{base}}} \frac{a}{r} 
   %                     + \left( \nabla a - \frac{a\vec{r}}{r^2} \right)\left( \frac{\vec{r}}{r} \cdot \nabla \frac{\partial \eta}{\partial \vec{r}_{\rm{base}}} \right)\\
   \vec{\nabla} \xi(\vec{r}) &= \vec{\nabla}_{\rm{base}}\xi(\vec{r}_{\rm{base}}) \frac ar 
                +\left(\vec{\nabla}_{\rm{base}}\xi(\vec{r}_{\rm{base}}) \cdot \hat{\vec r}\right) \left( \vec{\nabla} a -\frac ar \hat{\vec r}\right) \\
   \vec{\nabla} \eta(\vec{r}) &= \vec{\nabla}_{\rm{base}}\eta(\vec{r}_{\rm{base}}) \frac ar 
                +\left(\vec{\nabla}_{\rm{base}}\eta(\vec{r}_{\rm{base}}) \cdot \hat{\vec r}\right) \left( \vec{\nabla} a -\frac ar \hat{\vec r}\right) \\
   \vec{\nabla} \zeta(\vec{r}) &= \frac{2a}{r^2} \hat{\vec r} -\frac{2}{r}\vec{\nabla} a(\vec{r})
\end{align*}
%where $\vec{r}$ represents the cartesian coordinates, thus can be exchanged by $x$, $y$, or $z$, respectively.
%Whilefor first-order elements the base element is flat, higher-order elements may have a curved surface which has the form $a=\sum_i L_i(r) |\vec{r}_i|$ where $\vec{r}_i$ are the base elements' nodes and $L_i$ are the Lagrange functions used as basis.
In most cases, the base element must not be flat, but may have a curved surface which has the form $a=\sum_i \phi^{\rm{base}}_i(r) |\vec{r}_i|$ where $\vec{r}_i$ are the base elements' nodes and $\phi^{\rm{base}}_i$ are the basis functions of the base element.
Correspondingly, the gradient of $a$ reads
\begin{equation} \label{eq:grad_a}
   \vec{\nabla} a = \frac ar \frac{ \sum_i r_i \vec\nabla_{\rm{base}} \phi^{\rm{base}}_i(\vec{r}_{\rm{base}}) - \left(r_i \vec \nabla_{\rm{base}}\phi^{\rm{base}}_i(\vec{r}_{\rm{base}}) \cdot \hat{\vec r}\right) \hat{\vec r}}
                    {1-\sum_i r_i \left( \vec\nabla_{\rm{base}} \phi^{\rm{base}}_i(\vec{r}_{\rm{base}}) \cdot \hat{\vec r}\right) }
\end{equation}
where the gradients of the basis functions evaluated at the base element are, as mentioned above, assumed to be known.
The derivation of this formula and an expression for planar base elements are given in section \ref{sec:gradient}.
%The values $\frac{d\xi}{d\vec{r}_{\rm{base}}}$ and $\frac{d\eta}{d\vec{r}_{\rm{base}}}$ denote the mapping on the base face which, as mentioned above, is assumed to be known.

%\rt{Note: In case of a planar base we see that the equations are equivalent by noting that $\vec{r}_i$ lie on a plane and thus $\vec{r}_i\cdot\vec{n}$ all give the same number (following from the normal form of planes).
%Thus, the contribution from the last term vanishes because the gradients of the Lagrange polynomials sum up to zero which follows directly from the fact that a complete set of Lagrange polynomials sums to $1$ everywhere.}
%\bt{Note:While formally all terms in $\nabla a$ decay with $\frac 1r$, it changes direction due to the fact that $\vec{r}\cdot\vec{n}$ grows faster than linear with $\vec{r}$ because the two vectors have an increasing parallel contribution: While for $r =a$ the direction of $\nabla a$ is parallel to $n$, it points asymptotially in radial direction.}

\section{Setup of Matrices}
\label{sec:matrix}

As discussed, \textit{e.g.}, by Gerdes~\cite{Gerdes_CMAME_1998} and Burnett~\cite{Burnett_JASM_1994}, the setup of the system matrix is possible without divergence, but only when using a careful setup. % due to the divergence of the Jacobian. % discussed in~\sec{sec:problem}. %e.g. the mass matrix $\int |\phi(\vec{r})|^2 dV$ is not well-defined.
In the following, a modification to the matrix setup is discussed which provides such a setup, using analytic expressions on the physical element.
The starting point is the weak form of the differential operator
\begin{equation*}
   \int \phi_s \left(k^2+\Delta\right) \phi_t dV
\end{equation*}
where $\phi_s$ is a test function and $\phi_t$ is a trial function.
After application of the Gauss integration theorem, this gives the working expression
\begin{equation*}
   \int\left( -(\vec{\nabla} \phi_s)(\vec{\nabla} \phi_t)  + k^2 \phi_s \phi_t \right) J d\xi d\eta d\zeta
\end{equation*}
which is to be evaluated.

In general dimensional space, the Jacobian will grow, as discussed above, with $d^{-2}=r^{(\rm{dim}+1)}$ while solutions of the Helmholtz problem decay with $d\cdot r = r^{-\frac{\rm{dim}-1}{2}}$~\cite[ch. 5.4]{Leis_1986}, the derivatives $\vec{\nabla} \phi$ respectively with $d=r^{-\frac{\rm{dim}+1}{2}}$.
Thus, the mass matrix, containing only the Jacobian and the wave functions' square, diverges with $r^2$.
To get rid of this, different strategies are possible, depending on the type of infinite elements chosen.

\subsection{Astley-Leis elements}
%One way to resolve this is the use of Astley-Leis elements~\cite{Astley_JSV_1994}, where the test function is not identical to (the conjugate of) the trial function but has an extra $\frac{1}{r^2}$-term, leading to finite contributions for all terms separately.
In the scheme proposed by Astley \textit{et al.}~\cite{Astley_JSV_1994}, the test functions are written as
\begin{equation} \label{eq:basis}
   \phi_t(\vec{r})=\phi^{\rm{base}}_{t}(\vec{r}) \frac{e^{\i kr}}{r} p^n_t\left(\nicefrac 1r\right)
\end{equation}
where $\phi^{\rm{base}}_{t}$ is a shape function on the $\rm{dim}-1$-dimensional base element and $p^n(x)$ is an $n$-th order polynomial.
The index $t$ encodes the base element's test functions as well as the radial basis spanned by the polynomials $p$, which are independent degrees of freedom.

In contrast, the test functions have the form
\begin{equation*}
   \phi_s(\vec{r})=\phi^{\rm{base}}_{s}(\vec{r}) W(r) \frac{e^{-\i kr}}{r} p^n_s(\nicefrac 1r)
\end{equation*}
with the extra Sobolev weight $W(r)=\frac{1}{r^2}$.
%The complex conjugation is not necessary but has various advantages, for the following it is not important if it is taken into account or not.
Here the wave envelope form is used, but the unconjugated form works equivalently.

For setting up the matrix, the substitution
\begin{align*}
      J                 &\longrightarrow J\cdot d^2 \\
      \phi_{s,t}        &\longrightarrow \frac{\phi_{s,t}}{d} r \\
      \vec{\nabla} \phi_{s,t} &\longrightarrow \frac{\vec{\nabla} \phi_{s,t}}{d} r \\
      W                 &\longrightarrow \frac{W}{r^2} \\
      \vec{\nabla} W          &\longrightarrow \frac{\vec{\nabla} W}{r^2}
\end{align*}
results in all well-defined terms while the structure of the setup remains.
%Since the computation of these quantities generally must be specific for each element, this 

\subsection{Burnett elements}
%\begin{itemize}
%   \item asymptotic form of wave functions
%       in 3D: 1/r; in 2D: 1/sqrt(r); in 1D: 1
%   \item Mass matrix is infinitely large
%   \item full operator leads to finite matrix
%   \item Possible ways to setup the matrix
%   \item Do expressions represent polynomials? (or is a special integration necessary?)
%      Distinguish 'envelope' and 'unconjugated' elements.
%\end{itemize}
For Burnett elements~\cite{Burnett_JASM_1994}, the situation is more complicated since both $\phi_s$ and $\phi_t$ (\textit{i.e.} test and trial functions) are of the form \eq{eq:basis}, \textit{i.e.}, the extra weight $W$ is absent.

However, by restructuring the expressions the divergence can be eliminated.
The relevant change is based on the expansion of the gradients $\vec{\nabla} \phi_t=\i k\hat{\vec r} \phi_t +e^{\i kr} \vec{\nabla} \left(\phi^{\rm{base}}_{t}(\vec{r}) r^{-1} p^n_t(\nicefrac 1r) \right)$ for the test function and
$\vec{\nabla} \phi_s=-\i k\hat{\vec r} \phi_s +e^{-\i kr} \vec{\nabla} \left(\phi^{\rm{base}}_{s}(\vec{r}) r^{-1} p^n_s(\nicefrac 1r)\right)$ for the trial function, assuming the wave envelope form.
Insertion into the original weak form yields
%\begin{equation*}
\begin{multline*}
   \int\left[\left( -\i k\hat{\vec r} \phi_s +e^{-\i kr} \vec{\nabla} \left(\phi^{\rm{base}}_{s}(\vec{r}) \frac{1}{r} p^n_s(\nicefrac 1r)\right)  \right)
   \left(\i k\hat{\vec r} \phi_t +e^{\i kr} \vec\nabla \left(\phi^{\rm{base}}_{t}(\vec{r}) \frac{1}{r} p^n_t(\nicefrac 1r) \right) \right) - k^2 \phi_s \phi_t \right]
   dV \\
   = \int \left[\vec\nabla\left(  \phi^{\rm{base}}_{t}(\vec{r}) \frac{1}{r} p^n_t(\nicefrac 1r)\right) \vec\nabla\left( \phi^{\rm{base}}_{s}(\vec{r}) \frac{1}{r} p^n_s(\nicefrac 1r)\right) \right.\\
   % + \i k \phi_s \left(\i k\frac{\vec{r}}{r} \phi_t +e^{\i kr} \nabla \left(\phi^{\rm{base}}_{t}(\vec{r}) \frac{1}{r} p^n_t(\frac 1r) \right) \right) dV
   % - \i k \phi_t \left(\i k\frac{\vec{r}}{r} \phi_s +e^{-\i kr} \nabla \left(\phi^{\rm{base}}_{s}(\vec{r}) \frac{1}{r} p^n_s(\frac 1r) \right) \right) dV
   \left.+\left(\phi^{\rm{base}}_{s}(\vec{r}) \frac{1}{r} p^n_s(\nicefrac 1r)\right) \hat{\vec r} \vec\nabla \left(\phi^{\rm{base}}_{t}(\vec{r}) \frac{1}{r} p^n_t(\nicefrac 1r) \right)
   +\left(\phi^{\rm{base}}_{t}(\vec{r}) \frac{1}{r} p^n_t(\nicefrac 1r)\right) \hat{\vec r} \phi_s \vec\nabla \left(\phi^{\rm{base}}_{s}(\vec{r}) \frac{1}{r} p^n_s(\nicefrac 1r) \right) \right] dV
\end{multline*}
Finally, this form can be transformed to the reference space where the terms in brackets each decay with the power $\frac{1}{d}$ and thus cancel the divergence with $d^2$ from the Jacobian.

The disadvantage of this expression is that, being based on analytic cancellation of terms, the matrix assembly must follow an adapted strategy.
Since it involves derivatives of only some contributions to the basis functions, changes on many different places are required.
Note that, as discussed in the literature, in the case of Burnett elements the surface integral over the outer boundary does not vanish, leading to further contributions to the mass matrix due to application of the Gauss theorem.

\section{Supplement}
\label{sec:gradient}
\subsection{Gradient of $a$}

   When the base element is planar (\textit{e.g.}, is a first-order triangle), $a$ can be written as a projector of $\vec{r}$ onto $\vec r_{\rm{base}}$:
   The base points $\vec{r}_{\rm{base}}$ are described by the normal form of the plane $\vec{n}\left(\vec{r}_{\rm{base}}-\vec{r}_0\right)=0$ where $\vec{n}$ is the planes' normal vector and $\vec{r_0}$ is an arbitrary known point on the plane, \textit{e.g.} a node of the element.
Using this normal form, the function $a(\vec{r})$ reads
\begin{equation} \label{eq:a}
a(\vec{r})=\frac{\vec{r}_0 \cdot \vec{n}}{\vec{r}\cdot\vec{n}}r
\end{equation}
whose gradient has the form
\begin{equation} \label{eq:nabla_a1}
   \vec\nabla a = \frac{a}{r}\left(\hat{\vec r} - \frac{r \vec{n}}{\vec{r}\cdot\vec{n}}\right)\ .
\end{equation}

In the more general form, the base element is not planar but described via interpolation as 
\begin{equation*}
   a(\vec{r})= \sum_i |\vec{r}_i| \phi^{\rm{base}}_i(\vec{r}_{\rm{base}})
\end{equation*}
where usually $\phi^{\rm{base}}$ are Lagrange functions that are evaluated at the corresponding base point, see \eq{eq:rbase}.
Since the computation of the base point is easy in the reference space, these evaluation are straight-forward.
Computing the gradient of $a$ than leads to
\begin{align*}
   \vec{\nabla} a(\vec{r}) & = \sum_i |\vec{r}_i| \vec\nabla \phi^{\rm{base}}_i(\vec{r}_{\rm{base}}) \\
                    &=  \sum_i |\vec{r}_i| \frac{\partial \vec{r}_{\rm{base}}}{\partial \vec{r}} \frac{ \partial \phi^{\rm{base}}_i(\vec{r}_{\rm{base}})}{\partial \vec{r}_{\rm{base}}} \\
\end{align*}
   where  $\frac{\partial \vec{r}_{\rm{base}}}{\partial \vec{r}}$ is a tensor of partial derivatives, which is evaluated using $\vec{r_{\rm{base}}}=\frac{a}{r}\vec{r}$, leading to
\begin{equation} \label{eq:da_tensor}
   \frac{\partial \vec{r}_{\rm{base}}}{\partial \vec{r}} = \frac{\partial}{\partial \vec{r}}  \frac{a}{r}\vec{r} = \vec\nabla a \otimes \hat{\vec r} + \frac a r \left( 1 - \frac{\vec{r}\otimes \vec{r}}{r^2}\right).
\end{equation}
Inserting into the above expression, we obtain
   \begin{align*}
      \vec{\nabla} a(\vec{r}) & = \sum_i r_i \left(\vec\nabla a \otimes \hat{\vec r} + \frac a r \left( 1 - \hat{\vec r} \otimes \hat{\vec r} \right)\right) \vec\nabla_{\rm{base}} \phi^{\rm{base}}_i(\vec{r}_{\rm{base}}) \\
                    & = \sum_i r_i \left(\vec\nabla a \left(\hat{\vec r}\cdot \vec\nabla_{\rm{base}} \phi^{\rm{base}}_i(\vec{r}_{\rm{base}}) \right)
                                        + \frac ar  \vec\nabla_{\rm{base}} \phi^{\rm{base}}_i(\vec{r}_{\rm{base}})
                                        - \frac ar \hat{\vec r} \left(\vec\nabla_{\rm{base}}\phi^{\rm{base}}_i(\vec{r}_{\rm{base}}) \cdot \hat{\vec r} \right)
                                        \right) \\
                    %& = \frac ar \vec\nabla_{\rm{base}}\sum_i r_i \vec\nabla \phi^{\rm{base}}_i(\vec{r}_{\rm{base}}) - \left(r_i \vec\nabla_{\rm{base}} \phi^{\rm{base}}_i(\vec{r}_{\rm{base}}) \cdot \hat{\vec r}\right)\hat{\vec r}
                    %{1-\sum_i r_i \left( \vec\nabla_{\rm{base}} \phi^{\rm{base}}_i(\vec{r}_{\rm{base}}) \cdot \hat{\vec r}\right) }
                    & = \vec{\nabla} a \sum_i r_i \left(\hat{\vec r} \cdot \vec{\nabla}_{\rm{base}} \phi^{\rm{base}}_i(\vec{r}_{\rm{base}}) \right) \\
                    & \quad +\frac ar \sum_i \left( 
                    r_i \vec{\nabla}_{\rm{base}} \phi^{\rm{base}}_i(\vec{r}_{\rm{base}})
                    - r_i \hat{\vec r} \left( \vec{\nabla}_{\rm{base}} \phi^{\rm{base}}_i(\vec{r}_{\rm{base}}) \cdot \hat{\vec r}\right)
                    \right)
\end{align*}
   where $\vec\nabla_{\rm{base}} \phi^{\rm{base}}_i(\vec{r}_{\rm{base}})$ is the gradient of the base elements functions and thus known from the mapping of the base element.
Hence, from the last equation the working expression \eq{eq:grad_a} is obtained.

\subsection{Mapping of $\xi$ and $\eta$}

Since all of the following arguments hold equally for $\xi$ and $\eta$, the derivation is conducted for $\xi$ only.
The value of $\xi$ is obtained via the projected point $\vec{a}$ where the mapping is given. Thus, the gradient of $\xi(\vec{r})$ is obtained via the chain rule:
\begin{equation*}
  %\nabla \xi(\vec{r}) = \nabla \xi(\vec{a}) = \frac{\partial \xi}{\partial \vec{a}} \frac{\partial \vec{a}}{\partial \vec{r}}
   %\vec\nabla \xi(\vec{r}) = \vec\nabla \xi(\vec{a}) = \frac{\partial \xi}{\partial \vec{r}_{\rm{base}}} \frac{\partial \vec{r}_{\rm{base}}}{\partial \vec{r}}
   \vec\nabla \xi(\vec{r}) = \vec\nabla \xi(\vec{a}) = \vec{\nabla}_{\rm{base}} \xi(\vec{r}_{\rm{base}}) \frac{\partial \vec{r}_{\rm{base}}}{\partial \vec{r}}
\end{equation*}
where the second factor is again the tensor of partial derivatives, \eq{eq:da_tensor},
%where the first factor on the right $\frac{\partial \xi}{\partial \vec{a}}=\frac{d\xi}{d\vec{r}_{base}}$ and the second factor on the right hand side is a tensor which is, knowing the gradient of $a$, is eaiest to obtain by realising that $\vec{a}=\frac{a}{r}\vec{r}$ as \rt{This is used before; adjust formulation!}
%\begin{equation*}
  %\frac{\partial \vec{a}}{\partial \vec{r}} = \nabla a \otimes \frac{\vec{r}}{r} + \frac{a}{r} \left( \vec{\vec{1}} - \frac{\vec{r}\otimes\vec{r}}{r^2} \right)
%\end{equation*}
%where $\vec{\vec{1}}$ denotes the unity matrix.
%Thus, one obtains
which leads to
\begin{align*}
   \vec\nabla \xi(\vec{r}) & =  \left( \vec\nabla a \otimes \hat{\vec r} + \frac{a}{r} \left( \vec{\vec{1}} - \hat{\vec r}\otimes \hat{\vec r}\right)\right) \vec\nabla_{\rm{base}} \xi(\vec r_{\rm{base}}) \\
   & =  \left( \vec\nabla a \otimes \frac{\vec{r}}{r}\right) \cdot \vec\nabla_{\rm{base}} \xi(\vec r_{\rm{base}})%\frac{d\xi}{d\vec{r}_{\rm{base}}}
    + \frac{a}{r} \vec\nabla_{\rm{base}} \xi(\vec r_{\rm{base}}) %\frac{d\xi}{d\vec{r}_{\rm{base}}} 
    - \frac{a}{r} \frac{\vec{r}\otimes\vec{r}}{r^2} \vec\nabla_{\rm{base}} \xi(\vec r_{\rm{base}}) \\ % \frac{d\xi}{d\vec{r}_{\rm{base}}}  \\
   & = \vec\nabla a \left( \hat{\vec r} \cdot \vec\nabla_{\rm{base}} \xi(\vec r_{\rm{base}}) \right)  %\frac{d\xi}{d\vec{r}_{\rm{base}}}\right)
      + \frac{a}{r} \vec\nabla_{\rm{base}} \xi(\vec r_{\rm{base}}) %\frac{d\xi}{d\vec{r}_{\rm{base}}} 
    %- \frac{a\vec{r}}{r^2} \left(\hat{\vec r}\cdot \frac{d\xi}{d\vec{r}_{\rm{base}}} \right) \\
    - \frac{a\vec{r}}{r^2} \left(\hat{\vec r}\cdot  \vec\nabla_{\rm{base}} \xi(\vec r_{\rm{base}}) \right)
\end{align*}
which is the form specified above.

\newpage
\printbibliography

\end{document}